\documentclass[twocolumn,showpacs,preprintnumbers,amsmath,amssymb,prl]{revtex4-1}

\usepackage{graphicx}
\usepackage{dcolumn}
\usepackage{bm}

\begin{document}

\title{Anomalous magnetotransport in the heavy-fermion superconductor Ce$_2$PdIn$_8$}

\author{Daniel Gnida, Marcin Matusiak, and Dariusz Kaczorowski}
\affiliation{Institute of Low Temperature and Structure Research, Polish Academy of Sciences, P. O. Box 1410, 50-950 Wroc{\l}aw, Poland}

\date{\today}

\begin{abstract}
The normal state behavior in the heavy-fermion superconductor Ce$_2$PdIn$_8$ has been probed by means of Hall coefficient ($R_{\rm H}$) and transverse magnetoresistivity ($MR$) measurements. The results indicate the predominance of contributions from antiferromagnetic spin fluctuations at low temperatures. Anomalous non-Fermi-liquid-like features, observed below 8 K in both $R_{\rm H}(T)$ and $MR(T)$, are related to underlying quantum critical point, evidenced before in the specific heat and the electrical resistivity data. The magnetotransport in Ce$_2$PdIn$_8$ is shown to exhibit specific types of scaling that may appear universal for similar systems at the verge of magnetic instability.
\end{abstract}


\maketitle

In the vicinity of quantum critical point (QCP), which separates an ordered from disordered phase at zero temperature, the finite-temperature physical properties of heavy fermion (HF) systems are notably different from those predicted by the standard Fermi liquid (FL) theory of metals. As far as the magnetotransport properties are concerned, quantum criticality in two-dimensional (2D) antiferromagnetic (AF) systems manifests itself as a linear-in-$T$ electrical resistivity ($\rho $), strongly temperature- and field-dependent $R_{\rm H}$, and $MR$ obeying the so-called modified Kohler's rule \cite{kontani2008}.

There is still growing number of HF superconductors since the seminal discovery of unconventional superconductivity in CeCu$_2$Si$_2$ \cite{Pfleiderer}. Recently, a compound Ce$_2$PdIn$_8$ has been identified as an ambient-pressure HF superconductor with strongly enhanced normal-state electronic specific heat coefficient $\gamma \simeq$ 1.2 J/(mol K$^2$) and the transition temperature $T_{\rm c} \simeq $ 0.7 K \cite{218Pdb,218Pdbr,218Pdc}. The superconductivity in this material has been found to possess a nodal character \cite{218Pde}, likely of $d$-wave type, doubtlessly established for the closely related phase CeCoIn$_5$ \cite{An}. Furthermore, Ce$_2$PdIn$_8$ has been found to exhibit a magnetic-field induced QCP near the upper critical field $H_{c2}$(0) = 2.4 T \cite{218Pdd,218Pde,218Pdg}, again in strong similarity to CeCoIn$_5$ \cite{Paglione}. Most intriguingly, it has also been suggested that Ce$_2$PdIn$_8$ may fulfill the necessary conditions for the formation of a unique modulated Fulde-Ferrell-Larkin-Ovchinnikov superconducting state \cite{218Pde}, which possible occurrence at the magnetic high-field-low-temperature corner of the $H-T$ phase diagram of CeCoIn$_5$ is presently being hotly debated \cite{FFLO}. The many  common features between CeCoIn$_5$ and Ce$_2$PdIn$_8$ motivate further comprehensive studies of the physical behavior in the latter compound.

In this Communication we report on the Hall effect and the transverse magnetoresistivity of Ce$_2$PdIn$_8$, which both appeared highly anomalous yet consistent with the quantum criticality scenario, previously evoked based on the electrical resistivity \cite{218Pdc,218Pdd,218Pde}, thermoelectric \cite{218Pdf} and heat capacity \cite{218Pdbr,218Pdg} data. The polycrystalline sample used in the present study was previously characterized by means of X-ray diffraction, magnetic susceptibility, electrical resistivity and heat capacity measurements \cite{218Pdc}. Detailed metallurgical examination, in concert with the obtained physical property data, indicated single phase material, in particular free of any contamination by CeIn$_3$ impurity \cite{218Pdbr}. The magnetotransport measurements were performed on thin platelet-like polycrystalline specimen with the surface dimensions of 2.5 x 1.5 mm and thickness of 25 $\pm$ 5 $\mu \rm {m}$, which allowed recording the Hall voltage with sufficient precision even close to room temperature, where it was fairly small. The good reproducibility of Hall data was proved on another polycrystalline sample (cut from the same ingot) that was notably thicker (0.18 mm), yet had comparable surface dimensions (see the data in Ref.~\onlinecite{218Pdf}). The electrical leads were fabricated using silver paste to attach 50 $\mu \rm {m}$ thick silver wires to copper pads located on the sample surface. The pads were made by electrochemical deposition of copper from water solution of copper sulfate. This technique allowed to reduce the contact resistance down to about 0.1 $\Omega$. The experiments were carried out in the temperature interval 2 - 300 K and in applied magnetic field up to 9 T using a PPMS Quantum Design platform. The Hall effect was measured with flipping the position of the specimen in relation to the magnetic field direction using a horizontal rotator. This way the magnetoresistivity component could easily be separated from the Hall signal.

\begin{figure}
\includegraphics[width=\columnwidth]{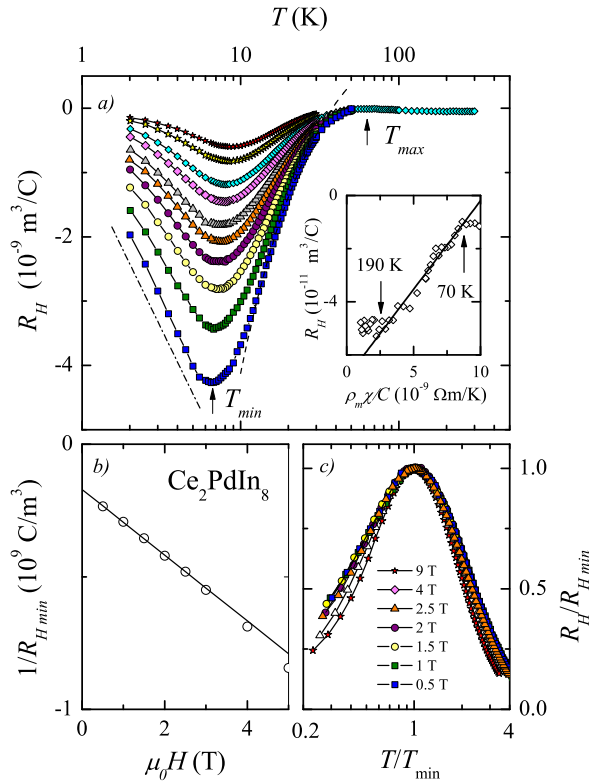}
\caption{\label{Fig1RHx} (Color online) (a) Temperature dependencies of the Hall coefficient of Ce$_2$PdIn$_8$ measured in several applied magnetic fields. The dashed and dash-dotted lines illustrate the behaviors discussed in the text. Inset: evaluation of the skew scattering contribution to $R_{\rm{H}}$ (see the text). (b) Field variation of $1/R_{\rm {Hmin}}$. The solid line emphasizes the linear dependence observed below 3 T. (c) Normalized Hall data versus normalized temperature (see the text for explanation).}
\end{figure}

As shown in Fig.\ref{Fig1RHx}a, the Hall coefficient of Ce$_{2}$PdIn$_{8}$ is negative in the entire temperature and magnetic field ranges studied. In the incoherent Kondo regime, i.e. above $T_{\rm {coh}}$ = 29 K \cite{218Pdc}, $R_{\rm {H}}$ is nearly independent of the temperature and the magnetic field strength. Its room-temperature value $R_{\rm{H}}^{300 \rm{K}}$ of $-4.8 \times 10^{-11}\rm{m}^{3}$/C is an order of magnitude smaller than $R_{\rm {H}}^{300 \rm{K}}$ measured for CeCoIn$_5$ \cite{Hundley115} and the Ce$_{2}T$In$_{8}$ phases with $T$ = Co, Rh, Ir \cite{218Co,218IrRh}. Within a single-band model one obtains an unrealistic concentration of electrons being $1.3 \times10^{29}$m$^{-3}$ (i.e. about $35 e^{-}/f.u.$). This finding indicates that Ce$_{2}$PdIn$_{8}$ is a compensated metal, in concert with the results of FS studies of the related compounds \cite{FS1,FS2,FS3,FS4}. Hence, the presence of multiple electron and hole bands must be taken into account in reliable calculations of carrier concentrations. In the overall Hall response above approximately 50 K,  one may identify a contribution due to skew scattering $R_{\rm{H}}^{\rm {skew}} = \xi \rho_{\rm {mag}} \chi /C$ (see the inset in Fig.\ref{Fig1RHx}a), where $\rho_{\rm {mag}}$ is the magnetic contribution to the electrical resistivity, $\chi$ is the molar magnetic susceptibility, and $C$ is the Curie constant (the data obtained for the very same polycrystalline specimen of Ce$_{2}$PdIn$_{8}$ as used in the present study were adopted from Ref.~\cite{218Pdc}). The coefficient $\xi$ represents the magnitude of the skew scattering effect and may be expressed (for $T \gg T_{\rm{K}}$) as $\xi=-\frac{5}{7} g \frac{\mu_{\rm {B}}}{k_{\rm {B}}}\sin{\delta}\cos{\delta}$, where $g$ stands for the magnetic ion Land\'{e} factor, $\mu_{\rm{B}}$ is the Bohr magneton, $k_{\rm {B}}$ is the Boltzmann constant and $\delta$ denotes the phase shift produced by incoherent Kondo scattering. The $\xi$ value derived for Ce$_{2}$PdIn$_{8}$ amounts to only 0.0067 K/T, which implies $\delta$ = -0.016 rad. This phase shift is somewhat smaller than those reported for CeCoIn$_5$ and the other Ce$_{2}T$In$_{8}$ phases \cite{Hundley115,218Co,218IrRh}. Clearly, in contrast to the generic behavior in heavy-fermion systems, the skew scattering mechanism in Ce$_{2}$PdIn$_{8}$ and the related compounds gives only a minor contribution to the overall Hall response. Although the microscopic origin of this anomalous feature remains unclear, the observed negligible magnitude of $R_{\rm{H}}^{\rm{skew}}$ allows a detailed analysis of the normal Hall effect.

In the coherent Kondo regime, the Hall coefficient of Ce$_{2}$PdIn$_{8}$ becomes strongly $T$- and $H$-dependent. In a field of 0.5 T, $R_{\rm{H}}$ exhibits a deep minimum at $T_{\rm {min}}$ = 6.5 K achieving a very large absolute value of $4.27 \times 10^{-9}$ m$^{3}$/C that is nearly two orders of magnitude larger than $R_{\rm{H}}^{300 \rm{K}}$. With increasing the magnetic field strength the minimum in $R_{\rm{H}}$($T$) broadens and slightly shifts to higher temperatures ($T_{\rm {min}} \simeq$ 8.5 K for $\mu_0 H = $ 9 T). Simultaneously, the magnitude of the Hall response at $T_{\rm {min}}$, denoted in the following as $R_{\rm {Hmin}}$, rapidly decreases, especially in weak magnetic fields, and follows the relationship $1/R_{\rm {Hmin}} \propto H$ for fields $\mu_{0} H < $ 3 T (see Fig.\ref{Fig1RHx}b). Interestingly, as can be inferred from Fig.\ref{Fig1RHx}c, plotting the experimental Hall data in the form $R_{\rm{H}}/R_{\rm {Hmin}}$ versus $T/T_{\rm {min}}$ one finds an universal behavior for $\mu_{0} H < $ 3 T, namely within the temperature interval 2 K $\leq T \leq T_{\rm{coh}}$ the $R_{\rm{H}}(T)$ dependencies collapse onto a single curve. In stronger fields one observes gradual deflection of $R_{\rm{H}}/R_{\rm {Hmin}}$ from the universal behavior on both sides of $T_{\rm {min}}$. Here, it is worth noting that $T_{\rm {min}}$ is very close to the temperature $T_{\rm {NFL}}$ below which the resistivity of Ce$_2$PdIn$_8$ varies linearly with the temperature \cite{218Pdc}, while the Seebeck $S/T$ and Nernst $\nu/T$ coefficients \cite{218Pdf}, as well as the electronic specific heat coefficient $C/T$ \cite{218Pdb,218Pdbr} exhibit logarithmic divergences. It seems that the observed low-temperature behavior of the Hall response is another indication of the non-Fermi-liquid state in Ce$_{2}$PdIn$_{8}$. Possibly, the derived scaling of the Hall response represents a generic feature of NFL systems, however this conjecture needs verification with other experimental data and calls for comprehensive theoretical input.

The overall temperature dependence of the Hall coefficient of Ce$_2$PdIn$_8$ is strikingly similar to that reported for CeCoIn$_5$ \cite{Hundley115}. For the latter compound, a theoretical approach has been developed \cite{kontani2008} that is based on the microscopic Fermi liquid theory with an important ingredient of the so-called current vertex correction (CVC). It appears that in nearly antiferromagnetic systems there occurs a strong backward scattering of charge carries due to critical magnetic fluctuations. This mechanism gives rise to an extra current of excited quasiparticles bearing significant momentum dependence, which manifests itself in distinct enhancements in the magnetotransport and thermoelectric effects. In particular, the Hall coefficient exhibits at low temperatures a huge increase in its absolute value, as observed for the Ce$T$In$_5$ ($T$ = Co, Rh) compounds but also for high-$T_c$ cuprates and some organic superconductors \cite{kontani2008}. Following the scenario formulated for CeCoIn$_5$ \cite{nkj2007}, the $R_{\rm{H}}$($T$) dependence measured for Ce$_2$PdIn$_8$ can be divided into three intervals: (i) high-temperature region, where the elastic scattering time is determined by isotropic phonon and incoherent Kondo scattering and hence $R_{\rm{H}}$ is almost $T$-independent, (ii) intermediate-$T$ region, for $T_{\rm {NFL}}$ = 8 K $< T <$ $T_{\rm {coh}}$ = 29 K, where the scattering time is highly anisotropic (because of the presence of hot and cold spots on the Fermi surface) and $R_{\rm{H}}$ is dominated by inelastic scattering due to antiferromagnetic fluctuations with strong CVC effect, and (iii) very low temperature region, in which the scattering time is governed by isotropic impurity scattering. The electronic transport theory developed by Kontani et al. \cite{kontani2008} assumed a 2D character of the Fermi surface and hence it predicted for the (ii) interval a $R_{\rm{H}} \propto \xi_{\rm{AF}}^2 \propto T^{-1}$ dependence ($\xi_{\rm{AF}}$ is the AF correlation length) , which has indeed been experimentally observed for CeCoIn$_5$ \cite{nkj2007}. In the case of Ce$_2$PdIn$_8$, this type of functional dependence of $R_{\rm{H}}$ appears less obvious. The relation predicted for a 2D AF system can be found in a rather narrow temperature range  12 K $< T <$ 30 K (see the dashed line in Fig.\ref{Fig1RHx}a), yet only if a more general formula $R_{\rm{H}} = R_{\rm{H}}^{0}+ A/(\theta + T)$ is applied. On the other hand, similarly good description of the Hall data in this temperature range can be obtained using a function $R_{\rm{H}} \propto A/(\theta + T)^{3/2}$, which has been found within the self-consistent renormalization theory for a purely three-dimensional AF system \cite{nkj2007}. Here it is worthwhile recalling that Ce$_2T$In$_8$ phases are believed to have less distinct 2D character than their Ce$T$In$_5$ counterparts, as their crystallographic unit cells consist of two 3D CeIn$_3$ units and a single 2D slab of $T$In$_2$ \cite{218Yashima}.

According to Ref. \onlinecite{nkj2007}, in CeCoIn$_5$ and related phases, the region of predominance of the anisotropic scattering ceases at the temperature below which $R_{\rm{H}}(T)$ becomes mainly governed by the isotropic impurity scattering. However, for NFL systems one should generally consider also a different scenario that accounts for temperature-dependent changes in the quasiparticles effective masses $m^{*}$. The logarithmic divergence of the electronic contribution to the specific heat, observed for Ce$_2$PdIn$_8$ below $T_{\rm {NFL}} \simeq$ 8 K \cite{218Pdc,218Pdg}, implies $m^{*} \sim C/T \sim \rm{ln} T$. Accordingly, one might expect that in a similar temperature region the Hall coefficient $R_{\rm{H}} \sim 1/m^{*}$ varies inversely proportional to $\rm{ln} T$. Interestingly, it seems that such a temperature dependence of $R_{\rm{H}}$ indeed occurs below $T_{\rm {min}} \simeq T_{\rm {NFL}}$ in weak magnetic field limit (see dash-dotted line in Fig.\ref{Fig1RHx}a). Recently, an approach called fermion condensation quantum phase transitions theory (FCQPT) has been derived for NFL systems, which explicitly takes into account $m^{*}$ dependencies on temperature, magnetic field and/or charge carrier density \cite{ShagPRL}. It seems tempting to numerically apply the FCQPT model to the magnetotransport data of Ce$_2$PdIn$_8$.

\begin{figure}
\includegraphics[width=\columnwidth]{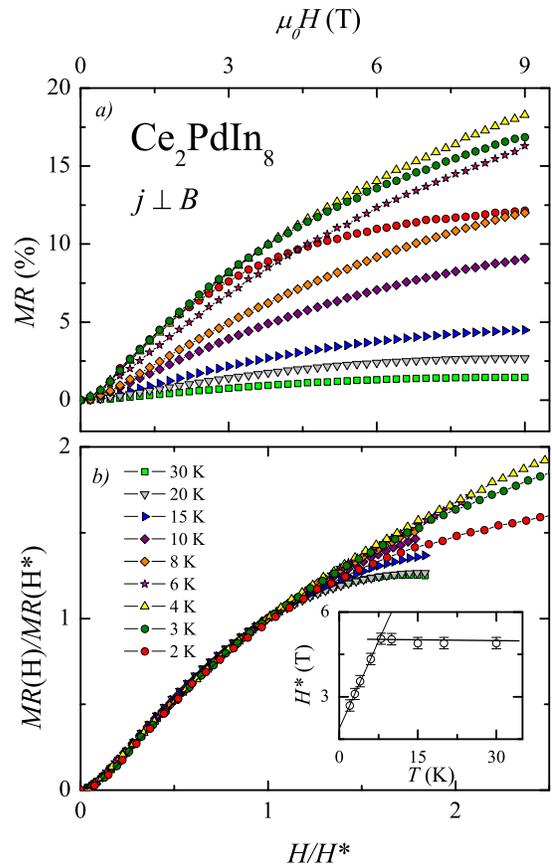}
\caption{\label{Fig2MRsc} (Color online) (a) Transverse magnetoresistivity isotherms of Ce$_2$PdIn$_8$ taken at a few different temperatures above 2 K. (b) Normalized magnetoresistivity data plotted versus normalized field (see the text). Inset: temperature dependence of the scaling parameter $H^*$.}
\end{figure}

\begin{figure}
\includegraphics[width=\columnwidth]{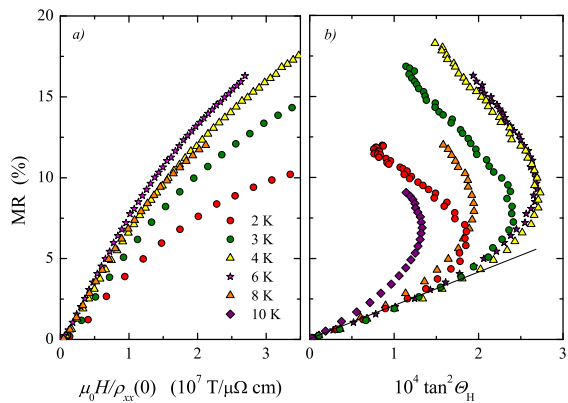}
\caption{\label{Fig3Kohler} (Color online) Scaling of the transverse magnetoresistivity of Ce$_2$PdIn$_8$ according to (a) the Kohler's rule and (b) the modified Kohler's rule (see the text for explanation). The straight line in (b) emphasizes a universal behavior.}
\end{figure}

Fig. \ref{Fig2MRsc}a presents the magnetic field variations of the transverse magnetoresistivity (defined as $MR(H) = [\rho_{xx}(H) - \rho_{xx}(0)]/\rho_{xx}(0)$) measured for Ce$_2$PdIn$_8$ in the temperature interval 2 - 30 K, i.e. in the coherent Kondo regime. In the entire temperature range studied, $MR$ is positive and its magnitude increases with increasing the field strength reaching about 18.2 $\%$ at 4 K in a field of 9 T. The observed convex curvature of the magnetoresistivity isotherms, the more evident the lower is the temperature, suggests development of a negative contribution in strong magnetic fields. This effect, likely arises from gradual suppression of spin-flip scattering by the magnetic field. One might expect that a broad maximum in $MR(H)$ would develop at temperatures lower than 2 K, with its position being shifted to smaller fields with rising the temperature, as observed e.g. for CeCoIn$_5$ \cite{nkj2007}. Interestingly, as shown in Fig. \ref{Fig2MRsc}b, the experimental $MR$ data can be plotted as the normalized magnetoresistivity $MR(H)/MR(H^*)$ versus the normalized field $H/H^*$ to get a universal curve over an extended interval in the magnetic field strength. The scaling parameter $H^*$, which provides an excellent overlap of all the isotherms, appears independent of the temperature above $T_{\rm {min}} \simeq T_{\rm {NFL}}$ = 8 K, and decreases linearly with $T$ at lower temperatures (see the inset to Fig.\ref{Fig2MRsc}b), i.e. in the range where the electrical resistivity \cite{218Pdc}, the specific heat \cite{218Pdb,218Pdbr}, the thermoelectric coefficients \cite{218Pdf} and the Hall coefficient of Ce$_2$PdIn$_8$ exhibit distinct non-Fermi liquid character. This coincidence suggests that the revealed behavior of $H^*(T)$ is another indication of the onset of quantum criticality in the compound studied. Hypothetically, such behavior of the magnetoresistivity might be universal for systems being at the verge of magnetic instability.

In the presence of strong backflow scattering of charge carriers on critical AF spin fluctuations, one can expect a violation of the Kohler's rule $MR = F(H/\rho_{xx})$ that is known to be well applicable to most FL systems \cite{kontani2001}. Instead, for CeCoIn$_5$ and other Ce$T$In$_5$ compounds with significant CVC contribution, a different scaling has been validated \cite{nkj2007}, namely $MR \propto {\rm{tan}}^2 \Theta_{\rm{H}}$, where $\Theta_{\rm{H}} = {\rm tan}^{-1}(\rho_{xy}/\rho_{xx})$ is the Hall angle. Remarkably, the latter so-called modified Kohler's rule (MKR) has also been experimentally observed for several high-$T_c$ superconductors \cite{kontani2001}. The results displayed in Fig.\ref{Fig3Kohler}a, clearly show that the Kohler's scaling does not work for Ce$_2$PdIn$_8$, however in some extended ranges of the parameters, the $MR$ data is proportional to ${\rm {tan}}^2 \Theta_{\rm{H}}$ (see Fig.\ref{Fig3Kohler}b). Moreover, in weak magnetic fields, the $MR$ isotherms measured in the NFL region (i.e. below 8 K) converge into a single curve that may approve for this compound the MKR approach.

In summary, in the coherent state, i.e. for $T_{\rm {coh}} <$ 29 K, the Hall coefficient of Ce$_2$PdIn$_8$ strongly depends on both the temperature and the magnetic field strength, while the transverse magnetoresistivity seems to obey the modified Kohler's rule, at least in the weak field limit. These anomalous magnetotransport properties exhibit close similarity to those reported before for CeCoIn$_{5}$ \cite{nkj2007}, and can be explained in terms of the recently developed theory in which anisotropy of the Fermi surface and strong backflow effect are taken into account \cite{kontani2001}. Undoubtedly, the low-temperature normal state electrical behavior of Ce$_2$PdIn$_8$ is governed by AF fluctuations, which eventually give rise to the onset of the unconventional superconductivity emerging near AF QCP \cite{218Pdd,218Pde, 218Pdg}. The derived scalings of the Hall effect data,  $R_{\rm{H}}(T)/R_{\rm{H}}$ vs. $T_{\rm{min}})$ and $R_{\rm{H min}}$ vs $H$, suggest the occurrence in the NFL state of a characteristic magnetic scale of the order of the upper critical field $H_{c2}$(0). Furthermore, in the extended temperature and magnetic field ranges, the normalized transverse  magnetoresistivity of Ce$_2$PdIn$_8$ has been found to exhibit a relationship $MR(H)/MR(H^*)$ vs. $H/H^*$ with a sole scaling parameter $H^*$ that is proportional to the temperature in the NFL state (below $T_{\rm {NFL}}$ = 8 K) and $T$- independent at higher temperatures (at least up to $T_{\rm {coh}} =$ 29 K). It remains to be verified, both experimentally and theoretically, if the magnetotransport scalings observed for Ce$_2$PdIn$_8$ may have universal applicability to heavy-fermion NFL systems being close to AF QCP instability.

The authors are grateful to V. H. Tran for fruitful discussions. This work was supported financially by the National Science Centre (Poland) under the research grant no. 2011/01/B/ST3/04482.

\bibliography{Ce2PdIn8b}



\end{document}